\documentclass[]{rmaa}

\usepackage{paralist}
\usepackage[T1]{fontenc}

\usepackage{psfrag,color}

\usepackage[latin1]{inputenc}
\usepackage{epsfig}
\usepackage{amsmath}
\usepackage{float}
\usepackage{graphicx}
\usepackage{subfig}
\usepackage{multirow}
\usepackage{array}
\usepackage{rotating}
\usepackage[normalem]{ulem}
\usepackage{wasysym}

\title{New spectroscopy of U\,Gem} 

\author{
J. Echevarr\'ia\altaffilmark{1},
S.H. Ram\'irez\altaffilmark{2},
M. Fuentes\altaffilmark{3},
L.J. S\'anchez\altaffilmark{1},\\
V. Pati\~no\altaffilmark{4,5} and
V. Chavushyan\altaffilmark{4}
}

\altaffiltext{1}{
Instituto de Astronom\'ia, 
Universidad Nacional Aut\'onoma de M\'exico, 
Apartado Postal 70-264, Ciudad Universitaria, 
M\'exico D.F., C.P. 04510, M\'exico}

\altaffiltext{2}{Department of Physics, University of Warwick, Coventry CV4 7AL, UK}

\altaffiltext{3}{Departamento de F\'isica, Universidad de Sonora, Unidad Regional Centro, Hermosillo, Blvd. Luis Encinas J, Calle Av. Rosales S/N, Centro, 83000, Sonora, M\'exico}

\altaffiltext{4}{
Instituto Nacional de Astrof\'isica, \'Optica y Electr\'onica, Luis Enrique Erro $\# 1$, Tonantzintla, Puebla 72840, M\'exico}
\altaffiltext{5}{Max-Planck-Institut f\"ur Radioastronomie, Auf dem H\"ugel 69, D-53121 Bonn, Germany
}

\shortauthor{Echevarr\'ia et al.}
\shorttitle{New observations of U\,Gem}

\abstract{We present new optical spectroscopic observations of U\,Geminorum obtained during a quiescent stage. We performed a radial velocity analysis of three Balmer emission lines yielding inconsistent results.
Assuming that the radial velocity semi amplitude accurately reflects the motion of the white dwarf, we arrive at masses for the primary which are in the range of $M_{\mathrm{wd}}= 1.21 - 1.37 M_{\astrosun}$. Based on the internal radial velocity inconsistencies and results produced from the Doppler tomography -- wherein we do not detect emission from the hot spot, but rather an intense asymmetric emission overlaying the disc, reminiscent of spiral arms -- we discuss the possibility that the overestimation of the masses  may be  due to variations of gas opacities and a partial truncation of the disc.

}

\resumen{Presentamos nuevas observaciones espectrosc\'opicas de U\,Geminorum obtenidas durante un estado de quietud. Realizamos un an\'alisis de velocidades radiales para tres lineas de Balmer, cuyos resultados son inconsistentes. Asumiendo que la semi-amplitud de la velocidad radial refleja fielmente el movimiento de la componente primaria, obtuvimos valores de masa de la primaria en el intervalo $M_{\mathrm{wd}}= 1.21 - 1.37 M_{\astrosun}$.
Basados en la inconsistencia de nuestros resultados y en las imagenes de tomograf\'ia Doppler -- donde no detectamos emisi\'on del punto caliente, sino que aparece una emisi\'on sobrepuesta al disco, que asemeja brazos espirales-- discutimos la posibilidad de que la sobreestimaci\'on de la masa sea debida a variaciones de las opacidades y un disco truncado.
}
\addkeyword{stars: cataclysmic variables, dwarf novae, individual (U\,Gem) -- techniques: spectroscopic}

\begin{document}

\maketitle

\section{Introduction}
\label{sec:intro}

U\,Geminorum is the prototype of a subclass of dwarf novae (DN), which belong to the Cataclysmic Variable systems (CVs). These are semi-detached interactive binaries where the primary is a compact white dwarf (WD) accreting material from a Roche-Lobe filling companion, which normally is a  late type star very close to the main-sequence. According to the classical model developed by  \citet{smak:1971} and \citet{warner:1971}, the material accreted by the secondary star forms an annulus or ring in the outer regions due to its large amount of angular momentum, and eventually forms a full disc, down to the boundary of the WD, due to viscous forces within its layers. When the disc is well formed the material strikes the disc in the outer rim which results in a conspicuous bright spot. This region, also known as the hot spot, is observed as an orbital hump in the optical light-curves of U\,Gem during quiescence, which precedes an eclipse of the bright spot and a partial eclipse of the accretion disc.\par

U\,Gem has an orbital period of 0.1769061911 days and a mass ratio of $q=0.35\pm 0.05$ \citep{echevarria:2007}, with an inclination of $i=69.7^{\circ}\pm0.7^{\circ}$ \citep{zhang:1987}. It has an outburst recurrence of $\sim118$ days. Models from \citet{takeo:2021} predict that the inner disc is truncated in quiescence at a distance of $\sim 1.20-1.25$ times the WD radius, whereas in outburst it truncates at $1.012$ $R_{\mathrm{wd}}$ or might even extend to the WD surface. The FUV lightcurve analyzed by \citet{godon:2017} shows phase-dependent modulations which are consistent with a stream overflow of the disc.\par
Multiple radial velocity studies have been conducted on U\,Gem, from which the semi-amplitudes of the components have been derived. Tracing the $H\alpha$ balmer emission line, \citet{echevarria:2007} obtained a radial velocity for the white dwarf of $K_1=107\pm2~\mathrm{kms^{-1}}$, in agreement with the analyses of FUV observations put forward by \citet{long:1999}, who reported a value of $K_1=107.1\pm2.1~\mathrm{kms}^{-1}$.\par
By means of Doppler Tomography -- a technique that analyzes the Doppler shifts of an emission line to obtain a two-dimensional distribution of the emission in accretion discs \citep{marsh:88}-- this object has been observed to exhibit diverse emission structures in quiescence: from that of an extended disc dominated by the emission from the hot spot \citep{echevarria:2007,marsh:1990}, to a highly asymmetric shape similar to spiral arms overlaying the disc \citep[e.g.][]{unda:2006,neustroev:1998}. \par
Despite being one of the best studied DN, and a prototype object, U\,Gem continues to show a behaviour far more complicated than that contemplated in the classical model. Thus, it is an object worth of continuous monitoring. With this in mind, in Section~\ref{sec:obs} we present optical spectroscopic observations of U\,Gem obtained during quiescence. Section~\ref{sec:radvel} is a radial velocity study of the systems implemented on three distinct emission lines: H$\beta$, H$\gamma$, and H$\delta$, by means of which the masses of the system are derived. Section~\ref{sec:discus} consists of the discussion of the derived masses. It also includes an extensive discussion on the Doppler tomography obtained for the emission lines, which we used to find clues on the spatial origin of the emission within the disc. Finally, our conclusions are presented in Section~\ref{sec:conclusions}.

\section{Observations and Reduction}
\label{sec:obs}
\begin{figure}
    \includegraphics[width=\columnwidth,angle=0]{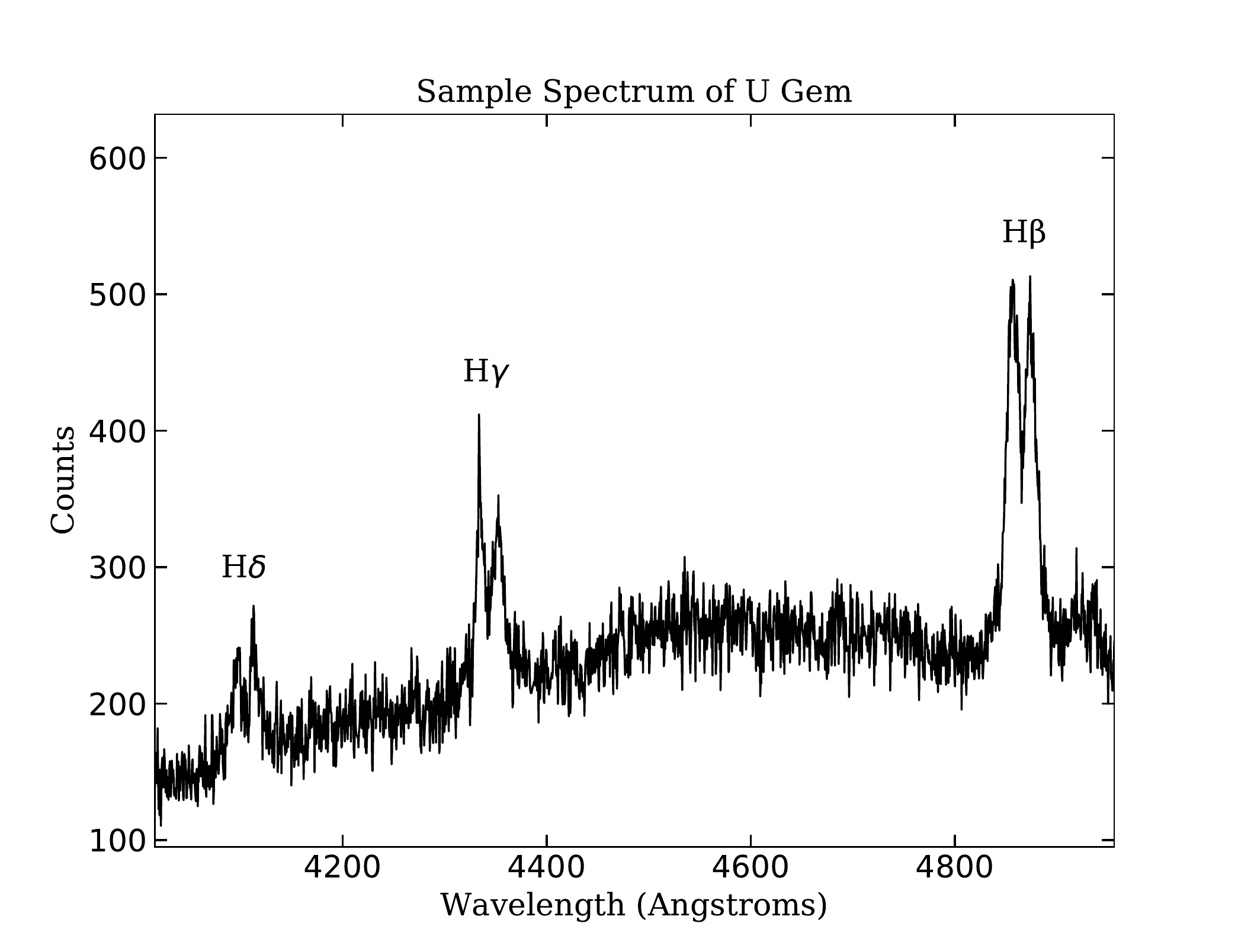}
    \caption{Individual spectrum of U\,Gem showing the strong double-peaked Balmer lines.}
    \label{fig:spec-exam}
\end{figure}

 Spectra were obtained with the 2.1-m telescope of the Observatorio Astrof\'isico Guillermo Haro at Cananea, Sonora, using the Boller and Chivens spectrograph and a E2V42-40, 2048x2048 CCD detector in the 4000 - 5000\,\AA \, range with a resolution of R~$\sim$ 1700, on the nights of 2021 February 15 and~16. The exposure time for each spectrum was 600\,s. Standard~{\sc iraf}\footnote{ IRAF is distributed by the National Optical Astronomy Observatories, which are operated by the Association of Universities for Research in Astronomy, Inc., under cooperative agreement with the National Science Foundation.}
procedures were used to reduce the data.   The log of observations is shown in Table~\ref{speclog}. The spectra show strong double-peaked Balmer lines, as exhibited in the sample in Figure~\ref{fig:spec-exam}. The spectra are not flux calibrated, therefore the y-axis shows counts in each spectrum. 
 
\begin{table}
\centering
	\caption{Log of spectroscopic observations of U\,Gem.}
    \label{speclog}
    \begin{tabular}{lccc}
        \hline
        Date & Julian Date & ~~~No. of     spectra &Exp.   \\
               & (2450000 +) &  &Time        \\
        \hline
15  February 2021 & 9260  & 47 & 600\,s \\
16  February 2021 & 9261  & 34  &600\,s \\

       \hline
\end{tabular}
\end{table}       

\section{Radial Velocities}
\label{sec:radvel}



The radial velocity of the emission lines of each spectrum was computed using the {\sc rvsao} package in {\sc iraf}, with the {\sc convrv} function, developed by J. Thorstensen (2008, private communication). This routine follows the algorithm described by \citet{schneider:1980}, convolving the emission line with an antisymmetric function, and assigning the centre of the line profile to the root of this convolution. As in \citet{segura:2020}, we used the 
 Double-Gaussian method ({\sc gau2} option available in the routine), which uses a negative and a positive Gaussian to convolve the emission line. The algorithm uses as input the width and separation of the Gaussians. This method traces the emission of the wings of the line profile, presumably arising from the inner parts of the accretion disc. \par
Following the methodology described by \citet{shafter:1986}, we made a diagnostic diagram to find the optimal Gaussian separation, by fitting each trial to a simple circular orbit :
\begin{equation}
V(t) = \gamma + K_1 \sin\left(2\pi\frac{t - t_0}{P_{orb}}\right),
\label{radvel}
\end{equation}
where $\gamma$ is the systemic velocity, $K_1$ the semi-amplitude (assumed to be the WD orbital velocity), $t_0$ the time of inferior conjunction of the donor and $P_{orb}$ is the orbital period. We employed $\chi^2$ as our goodness-of-fit parameter. Note that we have fixed the orbital period in our calculations, and therefore we only fit the other three orbital parameters. This is a convenient way to improve the fit of the remaining free  parameters as the orbital period has been obtained from the eclipses of the object \citep[e.g.][]{echevarria:2007}.

\begin{figure*} 
    \includegraphics[width=\columnwidth]{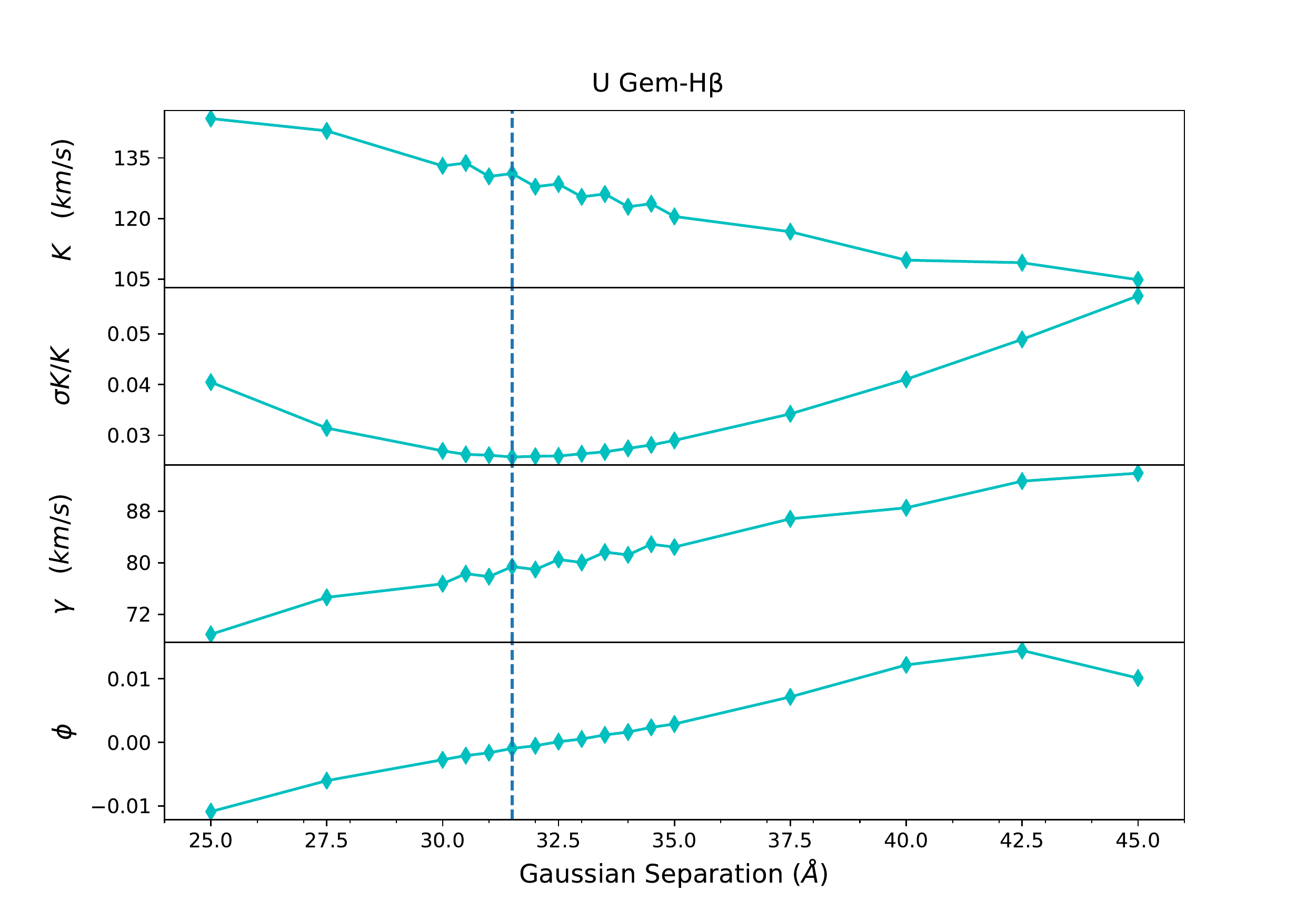}
    \caption{Diagnostic diagram of the $H\beta$ emission line. The vertical blue dashed line indicates the best solution. See text for further discussion.}
    \label{fig:diagnostic-hb}
\end{figure*}

\begin{figure*} 
    \includegraphics[width=\columnwidth]{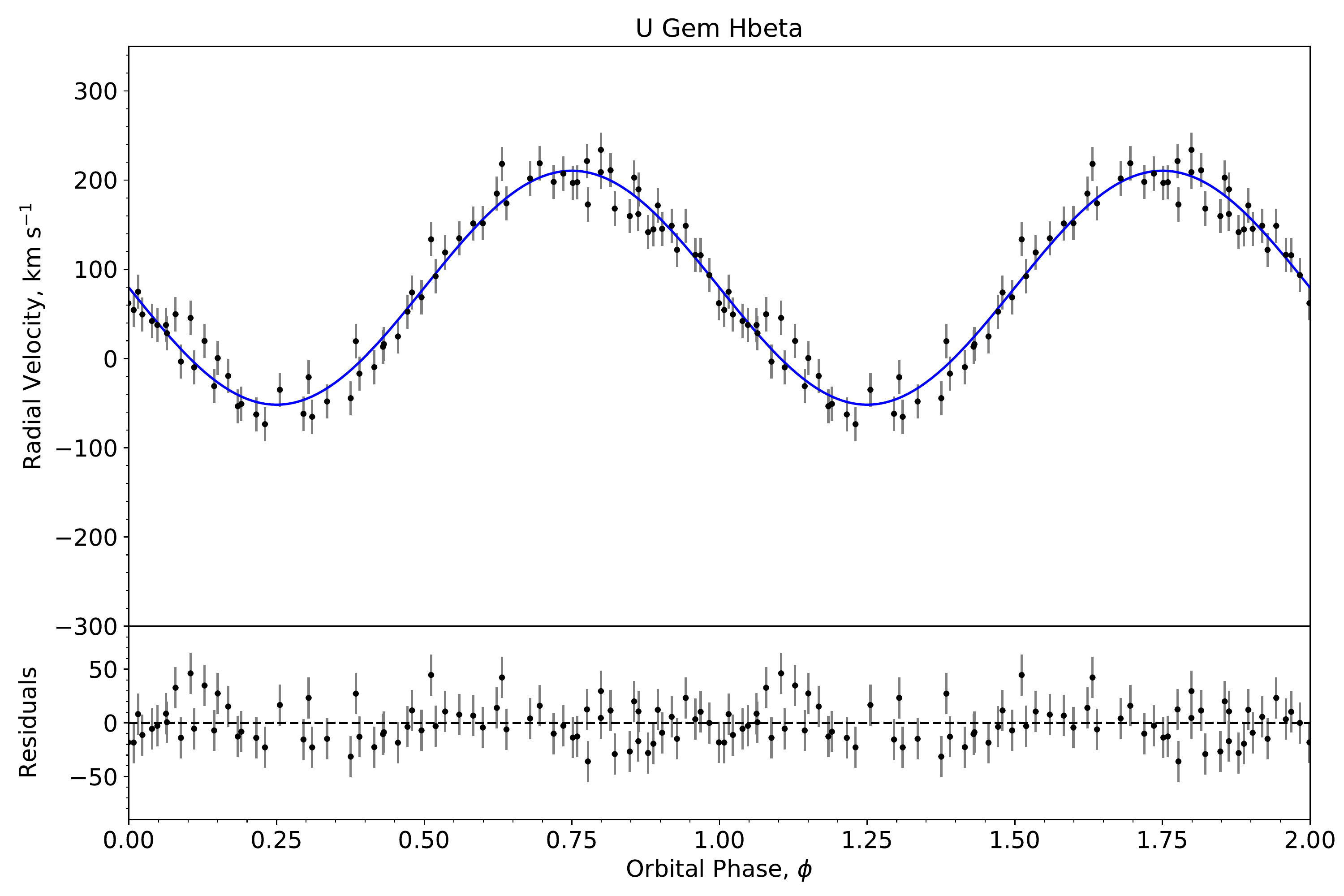}
    \caption{Radial velocity curve for the best solution of the $H\beta$ emission line.}
    \label{fig:radvel-hb}
\end{figure*}

Constructing a diagnostic diagram requires an interactive fitting between the {\sc convrv} routine and a program to fit the orbital parameters. We have used  {\sc orbital}\footnote{Available at \url{https://github.com/Alymantara/orbital_fit}}
a simple least square program to determine, in general, the three free orbital parameters. In particular, a control parameter is defined in this diagnostic, $\sigma_{K} / K$, whose minimum is a very good indicator of the optimal fit. In our runs we have found that the best results are obtained with a relatively small width of about 10--15 pixels.
The diagnostic diagram for $H\beta$ is displayed in Figure~\ref{fig:diagnostic-hb}, while the orbital fit for its best solution is exhibited in Figure~\ref{fig:radvel-hb}.
The parameters yielded for the optimal orbital fit are shown in Table~\ref{orbpar}. In a similar way, we have constructed the diagnostic diagrams for $H\gamma$ and $H\delta$. These and the corresponding best orbital fits are shown in Figures 4 to 7, while the orbital parametrs are also shown in Table~\ref{orbpar}.\par
The radial velocity fit for the $H\beta$ and $H\delta$ emission lines yield consistent $K_1$ values within the errors. They also agree with the value 
 of $K_1=138~\pm8~\mathrm{kms}^{-1}$, obtained from an analysis of the same lines performed by \citet{stover:1981}. On the other hand $H\gamma$ agrees with the more accurate result of $K_1=107.1\pm2.1~\mathrm{kms}^{-1}$ obtained by \citet{long:1999}, who traced the Doppler shifts of the 
 WD photospheric absorption lines  in the FUV range; also in agreement with \citet{echevarria:2007}, who followed the same methodology used  in this paper, but applied to the $H\alpha$ Balmer emission line only ($K_1=107\pm2~\mathrm{kms}^{-1}$). 

\begin{figure*} 
    \includegraphics[width=\columnwidth]{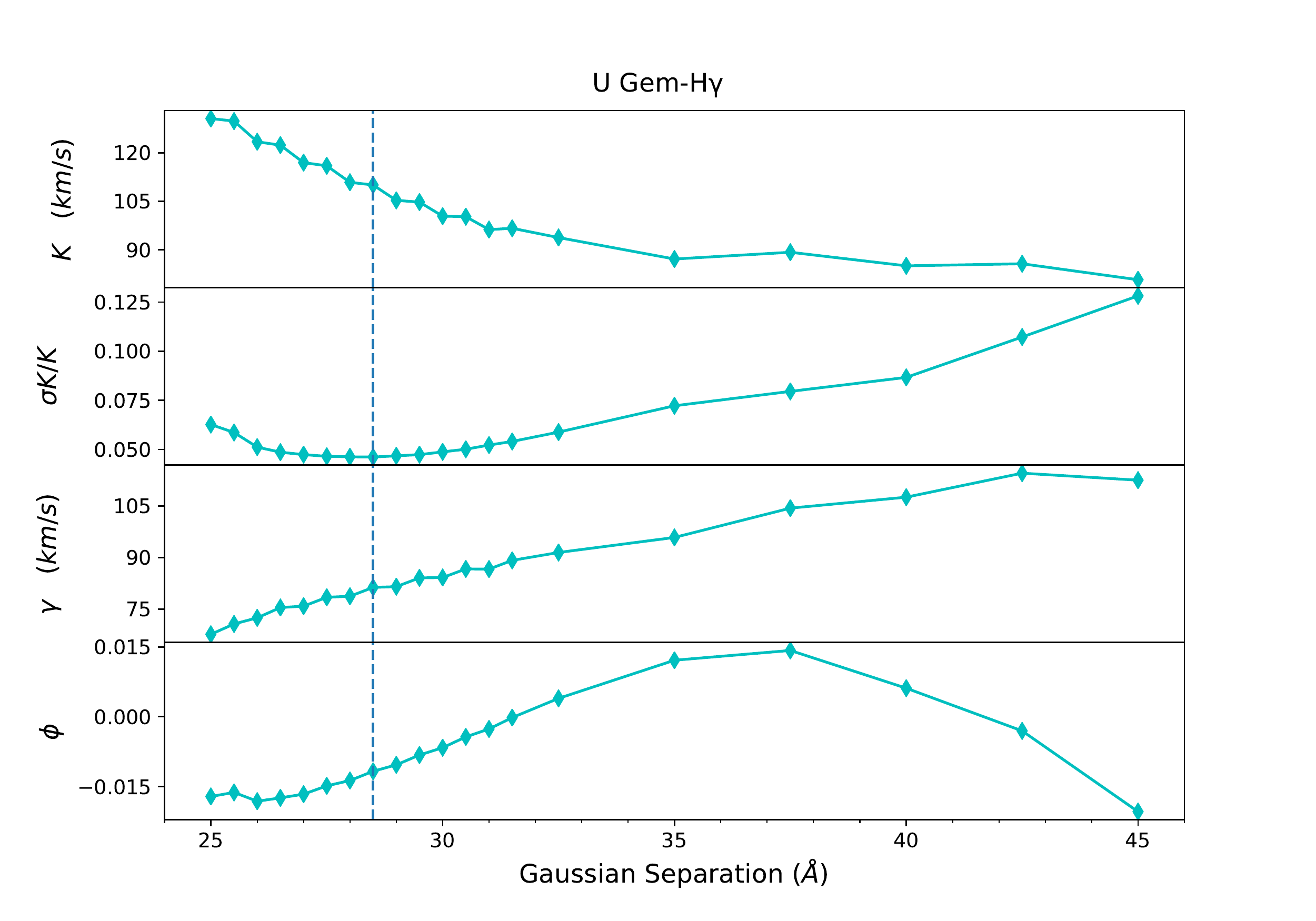}
    \caption{Diagnostic diagram of the $H\gamma$ emission line. The vertical blue dashed line indicates the best solution. See text for further discussion.}
    \label{fig:diagnostic-hg}
\end{figure*}

\begin{figure*} 
    \includegraphics[width=\columnwidth]{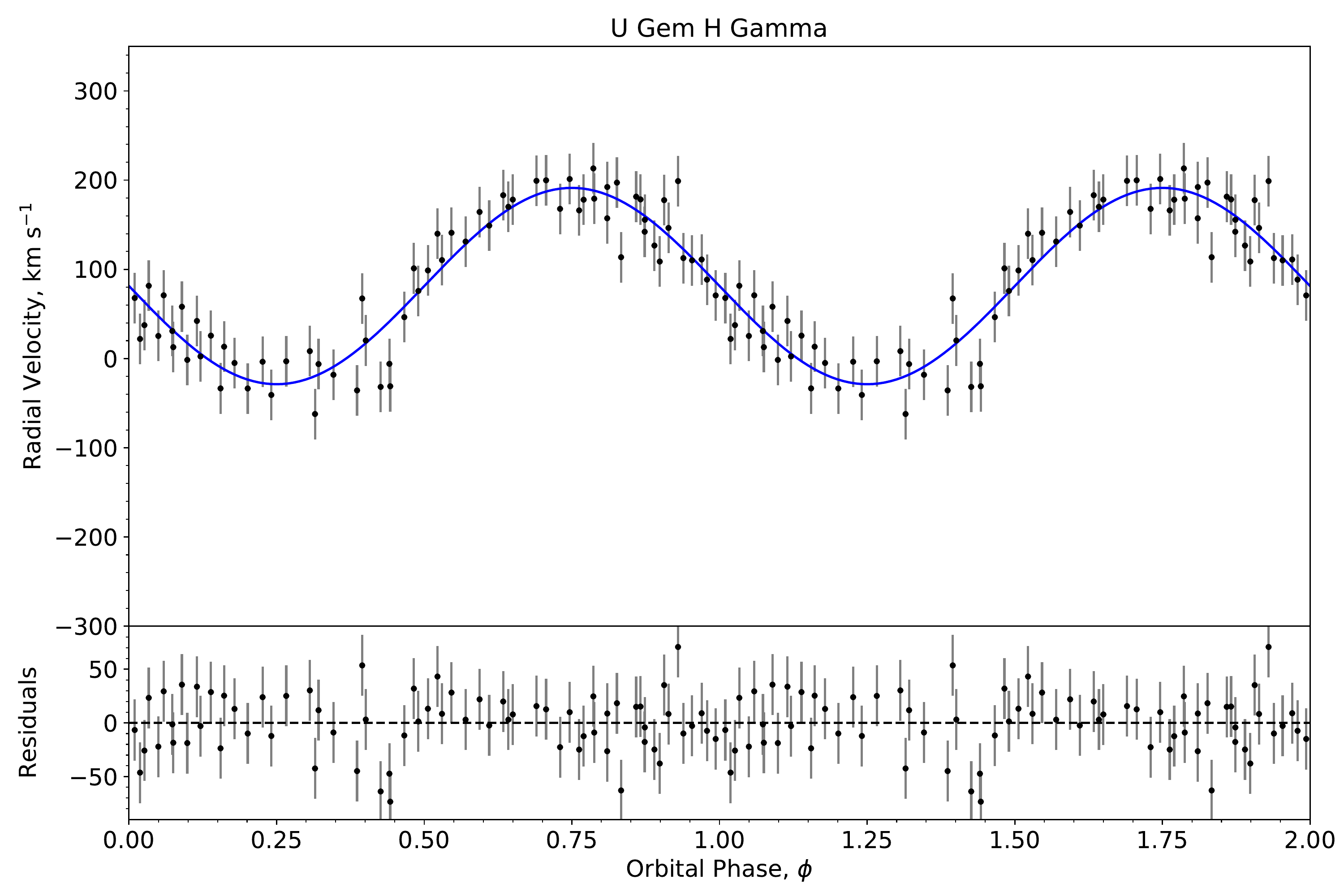}
    \caption{Radial velocity curve for the best solution of the $H\gamma$ emission line.}
    \label{fig:radvel-hg}
\end{figure*}

\begin{figure*} 
    \includegraphics[width=\columnwidth]{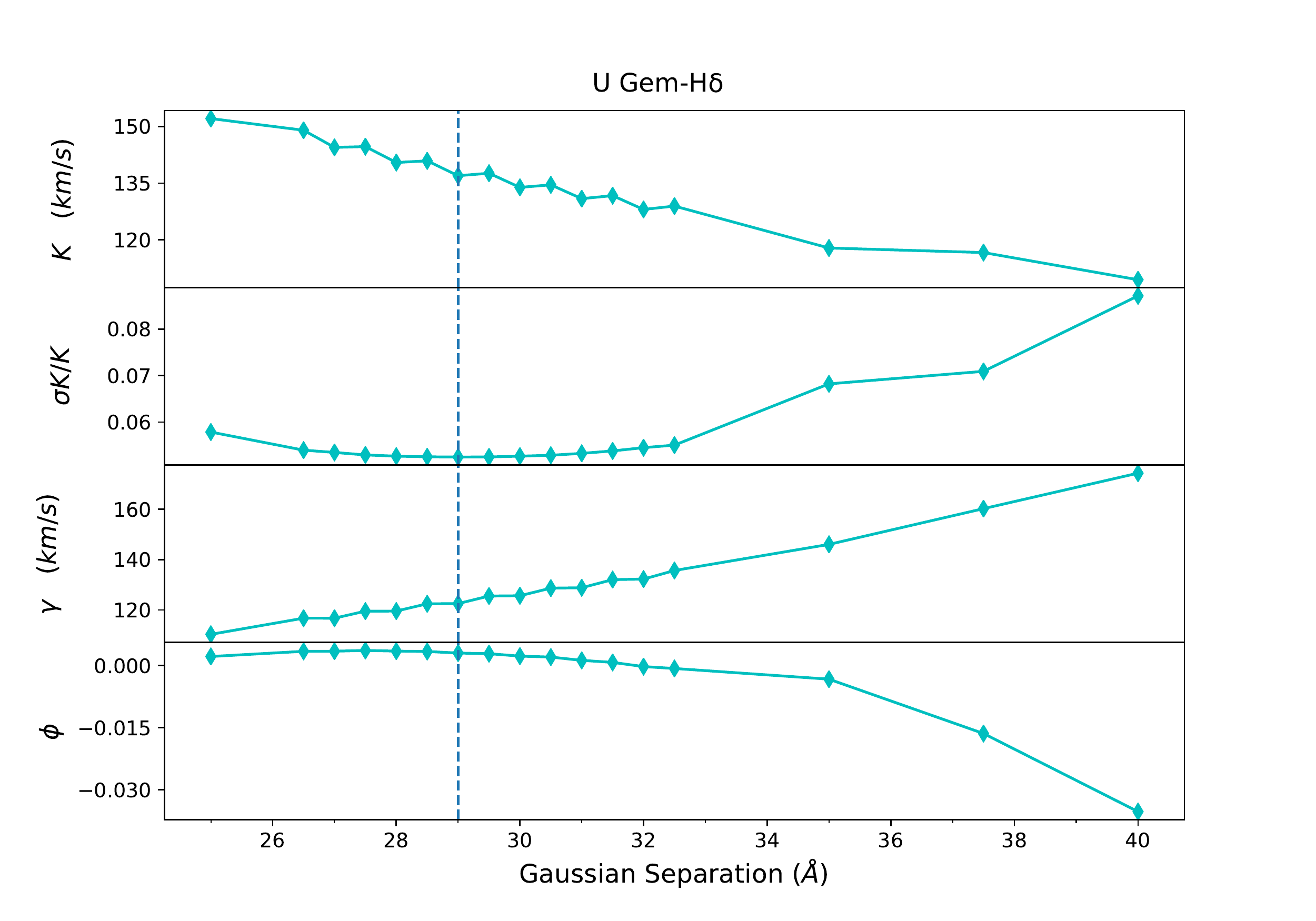}
    \caption{Diagnostic diagram of the $H\delta$ emission line. The vertical blue dashed line indicates the best solution. See text for further discussion.}
    \label{fig:diagnostic-hd}
\end{figure*}

\begin{figure*} 
    \includegraphics[width=\columnwidth]{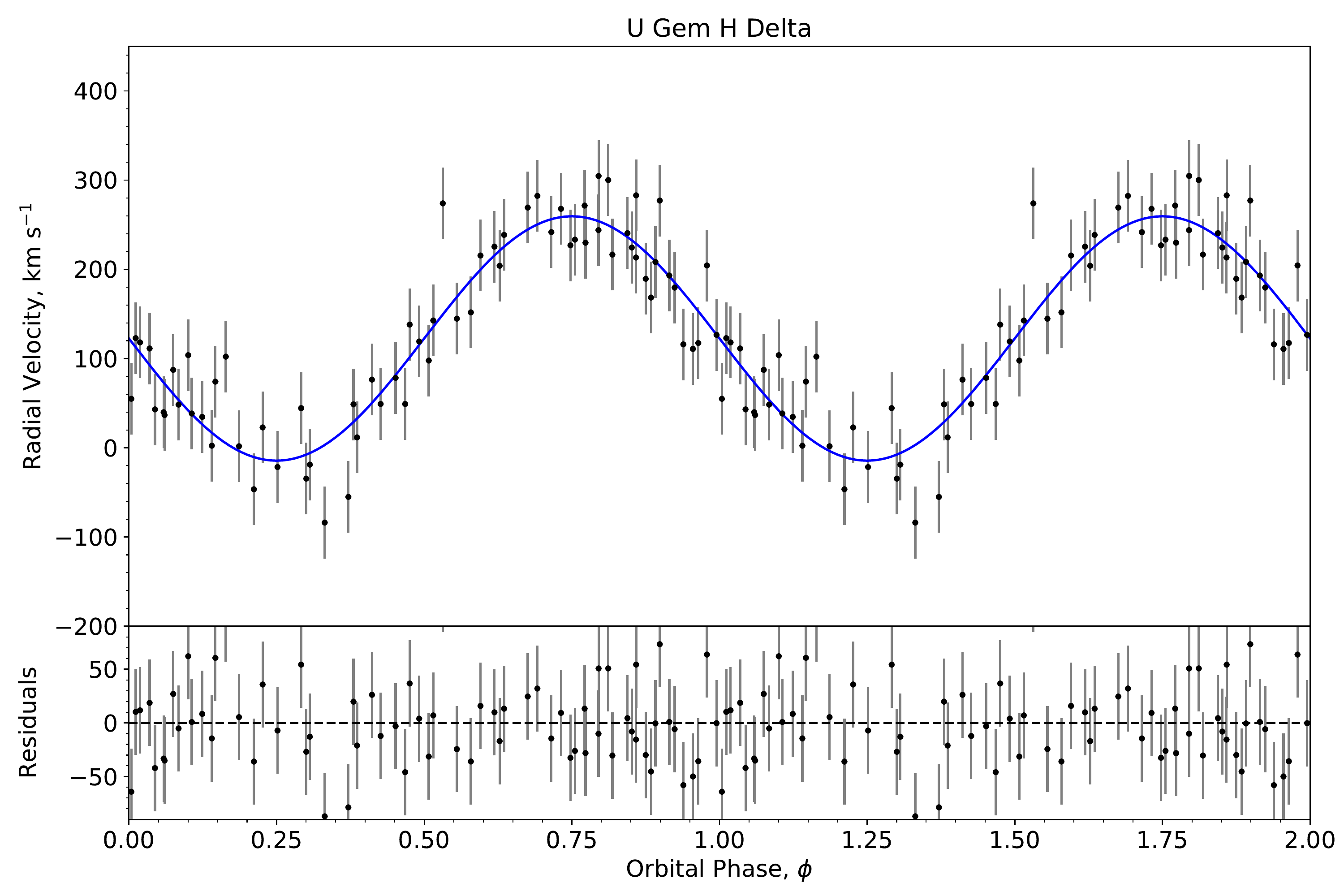}
    \caption{Radial velocity curve for the best solution of the $H\delta$ emission line.}
    \label{fig:radvel-hd}
\end{figure*}


\begin{table}

\centering
\caption{Orbital Parameters.} 
\label{orbpar}
\begin{tabular}{clll}
\hline
Parameter  &  H$\beta$ &H$\gamma$&H$\delta$                \\
\hline
  $\gamma$ (km\,s$^{-1}$) & 79.4 $\pm 2.3$&81.2$\pm 3.4$ & 122.5$\pm 4.9$     \\
   $K_1$ (km\,s$^{-1}$)   & 131.1 $\pm 3.3$&110.0$\pm 5.0$  &136.9 $\pm 7.1$     \\
  $HJD_0$*          & 0.0077 $\pm 0.0006$ &0.005$\pm 0.001$&0.008 $\pm 0.001$   \\
  $P_{orb}$  (min)        & Fixed**&Fixed** &Fixed** \\
\hline
\end{tabular}\\
*(2459261+ days)\\
**0.1769061911 days\\
\end{table}

\section{Discussion}
\label{sec:discus}
\subsection{Basic System Parameters}
\label{sec:bsp}
From the determination of the orbital parameters obtained in Section~\ref{sec:radvel} we can estimate the masses of the system components, as well as the binary separation; provided that an accurate estimation of the inclination angle is available. These mass estimates depend strongly on the assumption that the semi-amplitude derived from the emission lines accurately reflect the motion of the white dwarf, i.e. that the measurement of the wings of the lines are not distorted and present a symmetric behavior along the orbital period. The basic system parameters are obtained with the following formulae:
\begin{equation}
    \label{ec:q}
    q = \dfrac{K_{1}}{K_{2}} = \dfrac{M_{2}}{M_{1}}
\end{equation}
\begin{equation}
    \label{ec:M1}
    M_{1}sin^{3}i = \dfrac{PK_{2}(K_{1}+K_{2})^{2}}{2\pi G}
\end{equation}
\begin{equation}
    \label{ec:M2}
    M_{2}sin^{3}i = \dfrac{PK_{1}(K_{1}+K_{2})^{2}}{2\pi G} 
\end{equation}
\begin{equation}
\label{ec:a}
    a sin i= \dfrac{P(K_{1}+K_{2})}{2\pi} 
\end{equation}

Where $q$ is the mass ratio; $M_1$ is the mass of the primary; $M_2$ the mass of the secondary; $i$ the inclination angle, $K_1$ and $K_2$ are the semi-amplitude of the primary and secondary, respectively; and $a$ is the binary separation. To employ equations \ref{ec:q}--\ref{ec:a}, we adopted the inclination derived by \citet{zhang:1987} of $i=69.7^{\circ}\pm0.7^{\circ}$ and the semi-amplitude of the secondary derived by \citet{echevarria:2007} of $K_{2}= 310 \pm 5$ $\mathrm{kms^{-1}}$.

Table~\ref{sitpar} shows a summary for the system parameters yielded when using each of the $K_1$ values from the three lines, obtained in Section~\ref{sec:radvel}. For comparison, we also included the parameters reported for $H\alpha$ by \citet{echevarria:2007}.\\

\begin{table}[h]
\caption{Basic System Parameters yielded by the $K_1$ amplitude value of each emission line.}
\centering
\label{sitpar}
\begin{tabular}{lcccc}
\hline
Parameter  &$H\alpha^{\dag}$ & H$\beta$ &H$\gamma$&H$\delta$                \\
\hline
  q &$0.34\pm0.01$ &0.42 $\pm 0.01$ & 0.35 $\pm 0.01$ & 0.44 $\pm 0.02$ \\
  
  $M_{\mathrm{1}}$ ($M_{\astrosun}$) &$1.20\pm0.05$& 1.34 $\pm 0.05$ & 1.22 $\pm 0.06$ & 1.38 $\pm 0.07$ \\
  
  $M_{\mathrm{2}}$ ($M_{\astrosun}$) &$0.42\pm0.04$ &0.57 $\pm 0.02$ & 0.43 $\pm 0.03$ & 0.61 $\pm 0.05$ \\
  
  $a (R_{\astrosun})$ &$1.55\pm0.02$& 1.64 $\pm 0.02$ & 1.56 $\pm 0.03$ & 1.67 $\pm 0.03$ \\
\hline
\end{tabular}\\

\flushleft$^\dag$\citep{echevarria:2007}
\end{table}

As expected from the high $K_1$ values for H$\beta$ and H$\delta$, and because we are using the same $i$ and $K_2$ constraints as \citet{echevarria:2007}, the system parameters for these lines resulted in an overestimation with respect to those obtained from the $H\alpha$ analysis of the aforementioned authors  (See Table~\ref{sitpar}).
On the other hand, the parameters yielded for $H\gamma$ are consistent with those reported by \citet{echevarria:2007}, because of the agreement of the $K_1$ value.

A possible explanation for our radial velocity parameters overestimation and thus for our
mass parameter calculations could be made based on the X-ray analysis of \citet{takeo:2021}, whose models predict that the accretion disc is truncated at 1.25 Rwd during
quiescence, as expected by the theory \citep{narayan:1993}. 
Given that the Double-Gaussian method, employed in Section~\ref{sec:radvel}, traces the inner region of the disc, this truncation could result in higher values for the radial velocity of the WD.\par

It is possible that the the inner part of the disc does contain mass, but at such low density and low surface
brightness that it is optically thin \citep[e.g.][]{pringle:1981}. Furthermore, as explained in Section~\ref{sec:tom}, we detect an asymmetry overlaying the disc in our Doppler Tomograms. These circumstances could imply  abrupt variations of opacities within the disc, which would explain our internal inconsistencies in the radial velocity analysis \citep[e.g.][]{mason:2000}.

\subsection {Doppler Tomography}
\label{sec:tom}
\begin{figure}
\centering
	\includegraphics[angle=0,height=8cm,width=10cm]{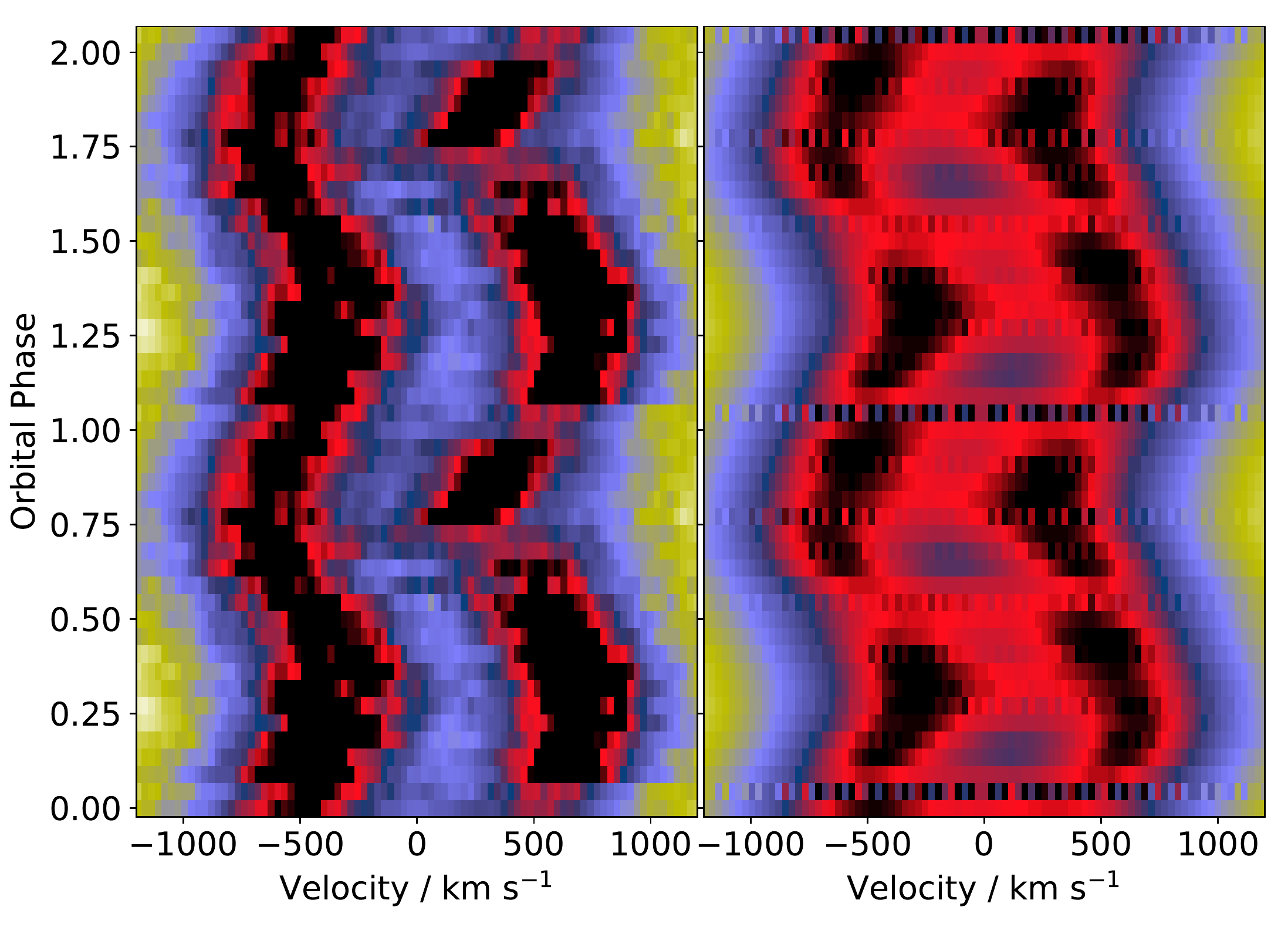}
	\includegraphics[angle=0,height=8cm,width=9 cm]{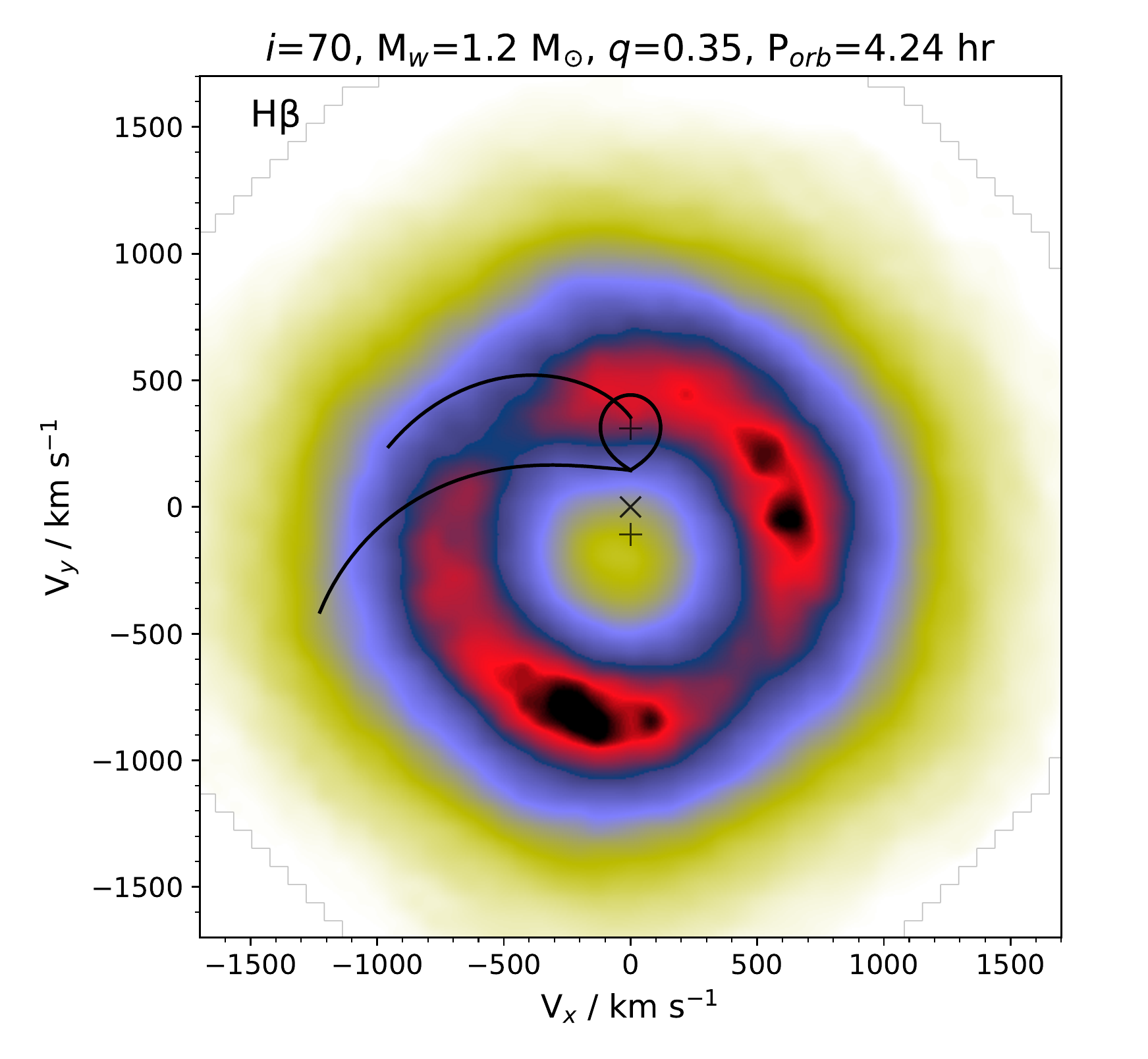}
\caption{Trail spectra and Doppler Tomography of the $H\beta$ emission line. The relative emission intensity is shown in a scale of colours, where the strongest intensity is represented by black, followed by red, then blue, and finally yellow. The cross markings represent (from top to bottom) the position of the secondary, the centre of mass and the primary component. The Roche lobe of the secondary is depicted around its cross. The Keplerian and ballistic trajectories of the gas stream are marked as the upper and lower
:  curves, respectively.}	
	
\label{dopmap-hb} 
\end{figure}

\begin{figure}
\centering
	\includegraphics[angle=0,height=8cm,width=10cm]{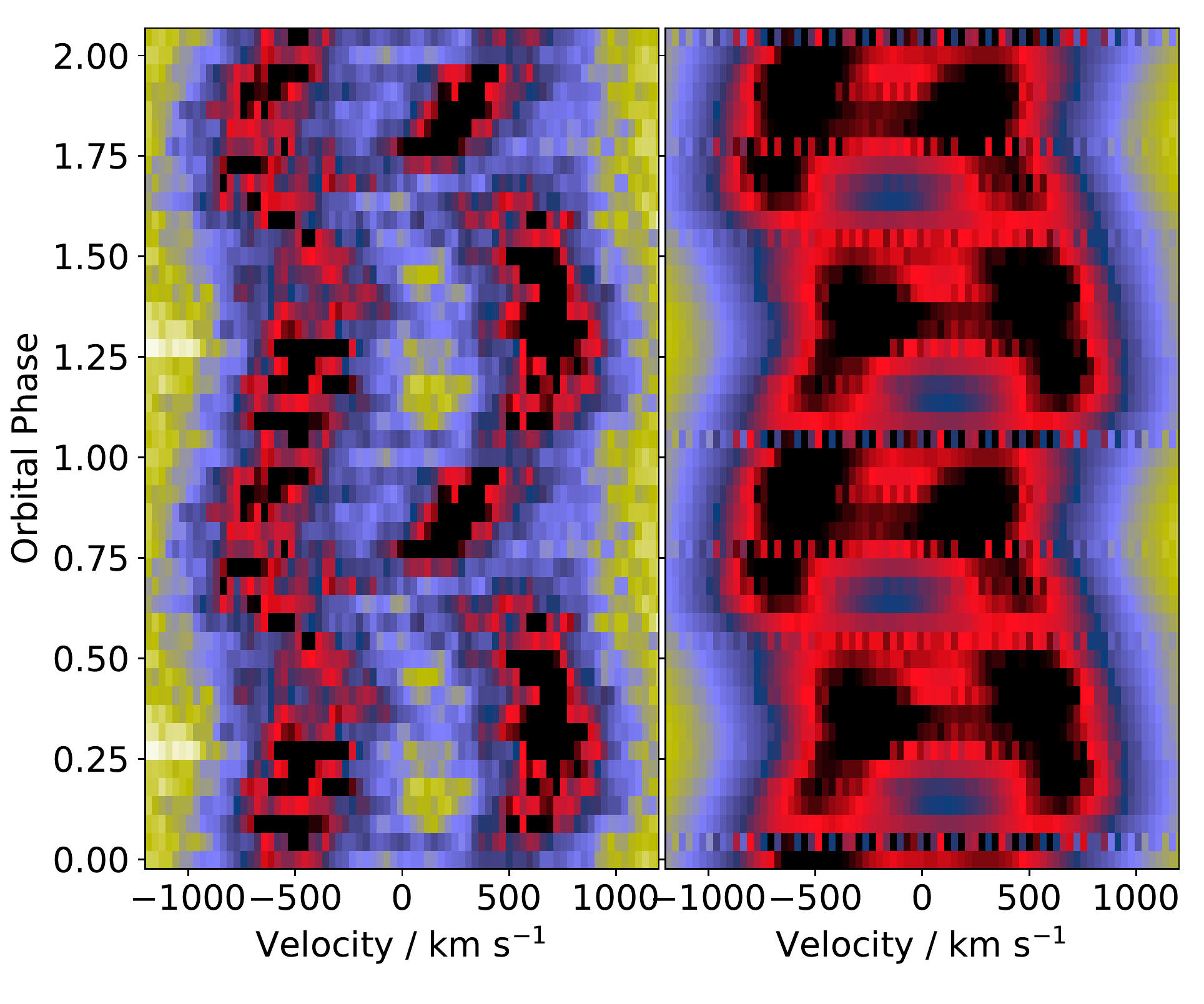}
	\includegraphics[angle=0,height=8cm,width=9 cm]{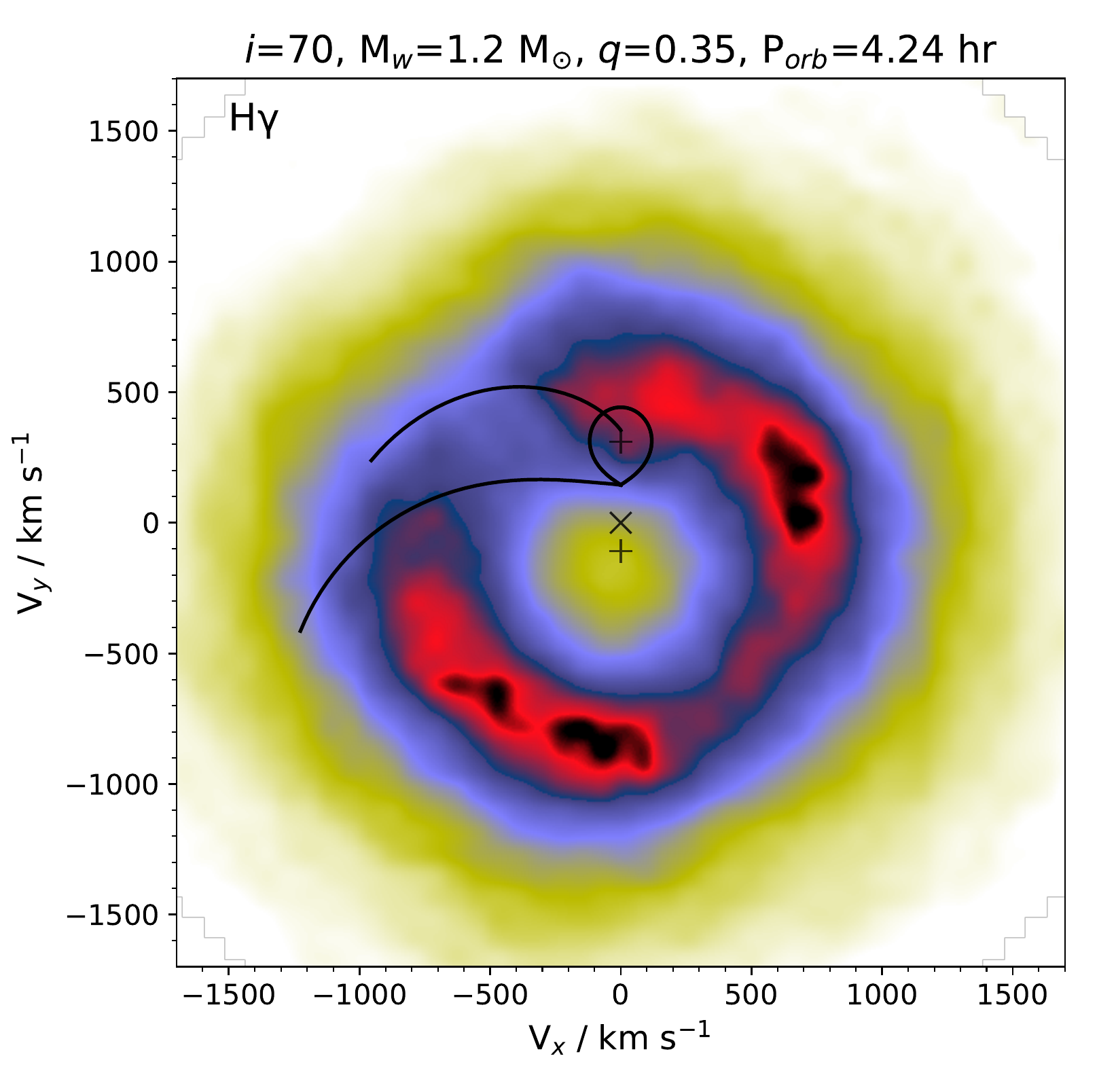}
\caption{Trail spectra and Doppler Tomography of the $H\gamma$ emission line. The relative emission intensity is shown in a scale of colours, where the strongest intensity is represented by black, followed by red, then blue, and finally yellow. The cross markings represent (from top to bottom) the position of the secondary, the centre of mass and the primary component. The Roche lobe of the secondary is depicted around its cross. The Keplerian and ballistic trajectories of the gas stream are marked as the upper and lower
curves, respectively.}	
	
\label{dopmap-hg} 
\end{figure}

\begin{figure}
\centering
	\includegraphics[angle=0,height=8cm,width=10cm]{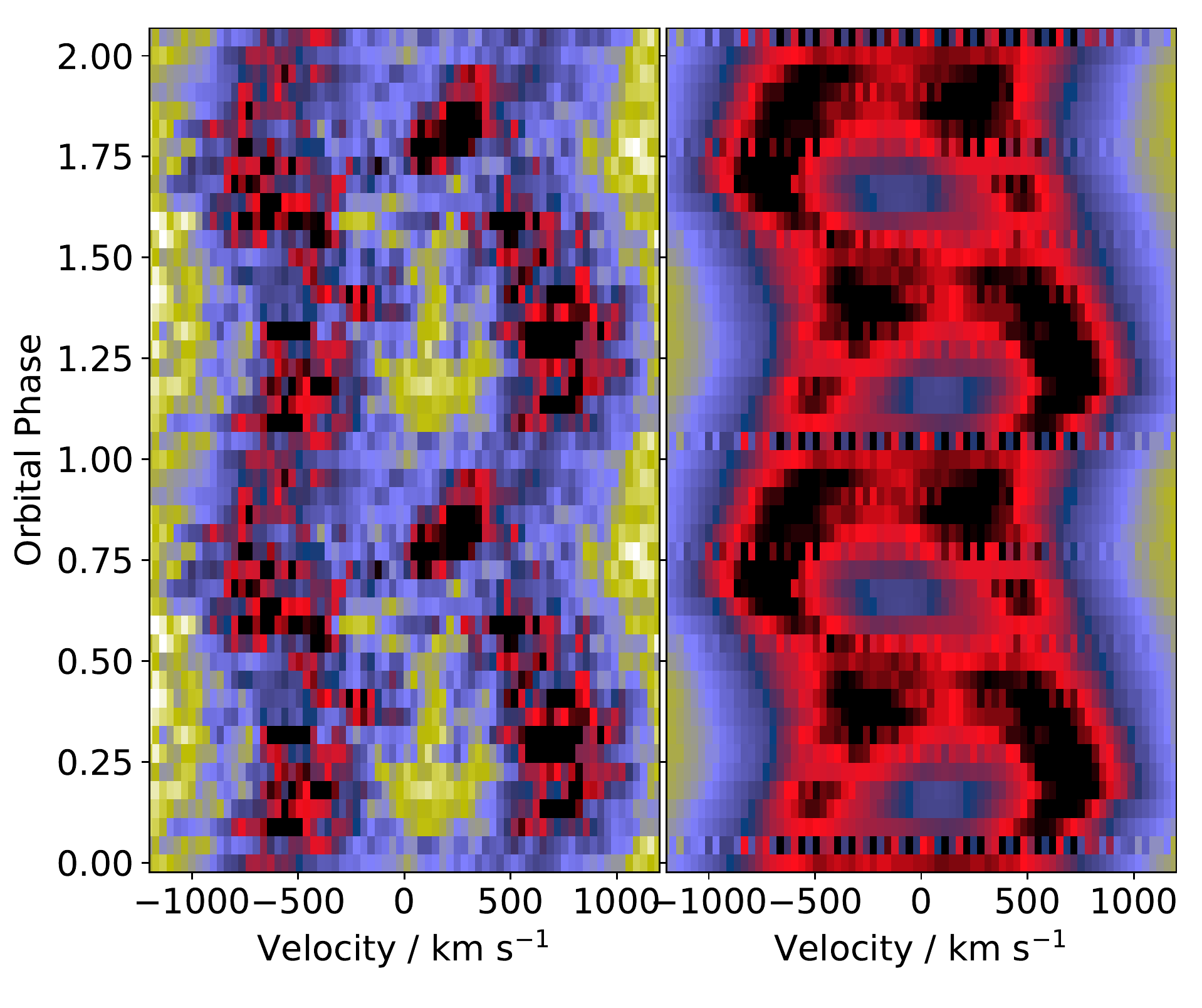}
	\includegraphics[angle=0,height=8cm,width=9 cm]{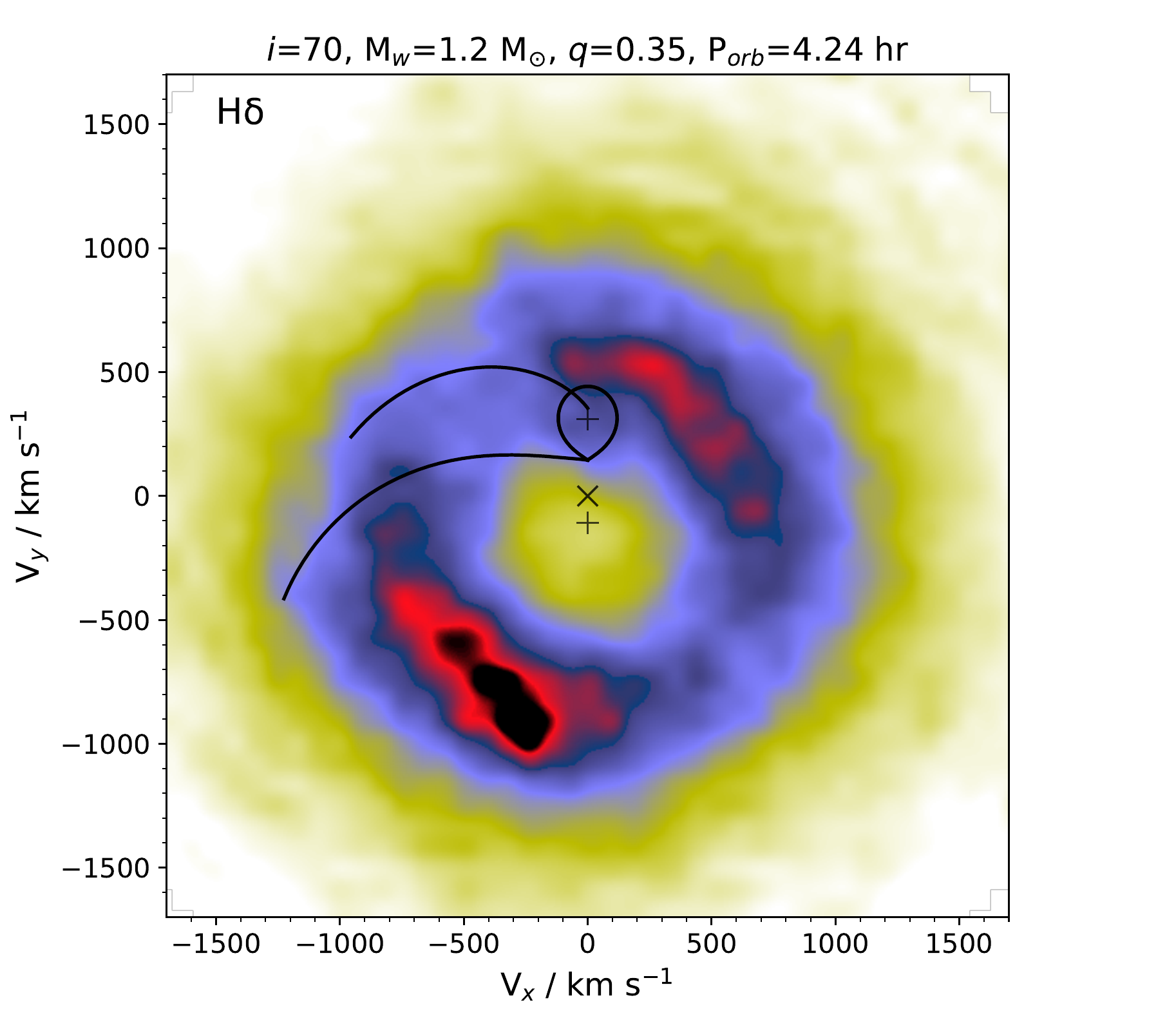}
\caption{Trail spectra and Doppler Tomography of the $H\delta$ emission line. The relative emission intensity is shown in a scale of colours, where the strongest intensity is represented by black, followed by red, then blue, and finally yellow. The cross markings represent (from top to bottom) the position of the secondary, the centre of mass and the primary component. The Roche lobe of the secondary is depicted around its cross. The Keplerian and ballistic trajectories of the gas stream are marked as the upper and lower
curves, respectively.}	
	
\label{dopmap-hd} 
\end{figure}

Doppler Tomography is an indirect imaging technique developed by \citet{marsh:88}. It produces two-dimensional mappings of the emission intensity in velocity space of the accretion disc, using the phase-resolved profiles of the spectral emission lines. We produced the Doppler Tomography of the $H\beta$, $H\gamma$ and $H\delta$ Balmer emission lines, using a Python wrapper \footnote{Available at \url{https://github.com/Alymantara/pydoppler}} \citep{pydoppler:2021} of the {\sc fortran} routines published by \citet{spruit:1998} within an {\sc idl} environment. 
Figures \ref{dopmap-hb}-\ref{dopmap-hd}, show the resulting images from the analysis, exhibiting the following layout:
In the top left panel we show the observed trailed spectra; the tomography is displayed in the bottom panel; and the reconstructed trailed spectra, which are created by collapsing the tomography image along the direction defined by the respective orbital phase \citep[][]{marsh:2005}, appear in the top right panel.

The trailed spectra of all three Balmer lines, show a conspicuous double-peaked structure, characteristic of the line profiles of discs in systems of high inclination \citep{horne:1986,marsh:88}. The spectrograms exhibit an evident lack of a hot-spot signature, which would appear as an s-wave oscillating from peak to peak.

The overall structure in our three tomography images is in contrast to most previous  Doppler tomography studies of U\,Gem in quiescence, which were dominated by an intense emission corresponding to a hot spot component \citep[e.g.][]{marsh:1990,echevarria:2007}. Instead, we find an asymmetric region of enhanced emission overlaying the disc, consistent with the structure exhibited by spiral density waves \citep[eg.][]{steeghs:1997}. While U\,Gem has been observed to show this structure in Doppler tomograms before \citep{neustroev:1998,unda:2006}, the presence of fully formed spiral shocks in U\,Gem must be regarded with caution: the study of the evolution of spiral shocks in U\,Gem performed by \citet{groot:2001} shows that they fade during the decline of the outburst. And even if spiral arms were present in U\,Gem in quiescence, they would be expected to be tightly wrapped during this stage \citep{steeghs:1999}; hence very difficult to detect with Doppler tomography \citep{ruiz:2020}. As discussed by \citet{unda:2006}, the spiral feature in the tomography could instead be explained by irradiation from the WD of regions of the disc that become thickened from tidal distortion. \par
Nonetheless, our understanding of spiral shocks from 2D models has shown limitations before, as the simulations by \citet{godon:1998} which predicted an unrealistically hot disc in order to reproduce the two-armed spiral pattern exhibited in the tomography of IP Peg by \citet{steeghs:1997}. Moreover, a previous detection in quiescence makes the limitations evident: the Doppler maps reported by \citet{pala:2019} exhibit a clear signature of spiral shocks in a quiescent state of the WZ\,Sge object SDSS\,J123813.73-033933.0; confirmed by the double hump modulation of the white dwarf in their \textit{HST} data, caused by the interface between the white dwarf and inner edge of the spiral shocks.  
And since the mass ratio of U\,Gem, $q=0.35\pm0.03$ \citep{echevarria:2007}, sets it right on the limit that allows the disc to achieve the 3:1 resonance ($q\lesssim0.3$) \citep{hellier:2001}, spiral arms can not be completely discarded. 

Another possible explanation is that provided by \citet{smak:2001}, which argues that the high radial velocity of $K_{\mathrm{1}}=138\pm8$~$\mathrm{kms^{-1}}$  obtained by \citet{stover:1981} (in accordance with our own values yielded for $H\beta$ and $H\delta$) could be caused by stream overflow of the disc. This would explain the absence of a hotspot in our tomography images, since the stream overshoot would avoid (or ameliorate) the initial impact with the rim of the disc. Stream material overshooting the disc edge and re-impacting at radii with lower velocity can create a second hotspot \citep{lubow:1989} which usually shows up in regions within the lower quadrants of the Doppler tomograms, as is the case in SW Sextantis systems \citep{schmidtobreick:2017}. This scenario is further supported by the phase-dependent modulation of the FUV light
curve and absorption lines velocity reported by \citet{froning:2001}, which
can be explained by the stream overflowing the edge of the disc \citep{godon:2017,godon:2019}.  \par

In any case, it is puzzling that our Tomography shows similar emission distributions for all three emission lines, implying that they arise from the same regions in the disc. This raises the question: why is it that for $H\beta$ and $H\delta$ we obtain values that are consistent with \citet{stover:1981}, which are likely corrupted by some additional effect on the disc; while on the other hand $H\gamma$ appears unaffected and agrees better with the more reliable WD radial velocity measured from \textit{HST} FUV observations by \citet{long:1999}? As mentioned in Section~\ref{sec:bsp}, we expect this internal inconsistency to be caused by different gas opacities in the accretion disc \citep{mason:2000}, occurring as a consequence of some combination of the phenomena discussed above: WD irradiation of tidally thickened regions, stream overflow, a partially truncated disc, and perhaps even fully formed spiral arms. \par
However, this inconsistency is a clear example of the issues arising from measuring the radial velocity of the WD from optical data, even when presumably tracing the inner regions of the disc as is done in the Double-Gaussian method (See Section~\ref{sec:radvel}).

\section{Conclusions}

\label{sec:conclusions}
We presented a spectroscopic analysis of the dwarf nova U\,Geminorum. We obtained the radial velocity of the system for three distinct Balmer emission lines: H$\beta$, H$\gamma$, and H$\delta$, by tracing the outer regions of the profile (which arise from the inner sections of the accretion disc), with the purpose of obtaining the WD radial velocity $K_1$. The resulting semi-amplitude for H$\gamma$ is consistent with previous canonic results of $K_1=107.1\pm2.1~\mathrm{kms}^{-1}$ 
 \citep{echevarria:2007,long:1999}, however, the other two lines show a considerable discrepancy, agreeing instead with the value obtained by \citet{stover:1981} of $K_1=138\pm8~\mathrm{kms}^{-1}$. We expected to find the source of this inconsistency in the Doppler tomography study, but the Tomograms show that the origin of all three lines arise from the same region. However, it must be noted that the tomography does not show a typical disc:  in particular there is no evidence whatsoever of a hotspot; instead it exhibits a spiral arm structure unexpected of a system in quiescence. 
This unusual shape (which can be a product of stream overflow, WD irradiation or actual spiral arms), along with a partial truncation of the inner regions of the disc, could together amount to considerable differences of gas opacities within the accretion disc, which could explain the different values of $K_1$ obtained for our three emission lines \citep[e.g.][]{mason:2000}.   \par
U\,Gem stands as one of the best studied DN, however as it is made evident in this paper,  more ingredients than those prescribed by the classical model must come into play to better explain its behaviour. Therefore, we propose further observations of this source to help shed light on the mechanisms giving rise to its rich and interesting nature.

\section*{Acknowledgements}

The authors are indebted to DGAPA (Universidad Nacional Aut\'onoma de M\'exico) support, PAPIIT projects IN114917 and IN103120. This project has received funding from the European Research Council (ERC) under the European Union's Horizon 2020 research and innovation programme (Grant agreement No. 101020057). This work was supported by CONACyT (Consejo Nacional de Ciencia y Tecnolog\'ia) research grants 280789 and 320987. This work was supported by the MPlfR-Mexico Max Planck Partner group led by V.M.P.-A. We thank the staff at the Observatorio Astrof\'isico Guillermo Haro  for facilitating and helping us to obtain our observations. This research made extensive use of {\sc astropy}, a community-developed core Python package for Astronomy \citep{Astropy-Collaboration:2013aa}, Python's SciPy signal processing library \citep[][]{virtanen:2020}, and {\sc matplotlib} \citep{Hunter:2007aa}.

\end{document}